\documentclass[aps,pre,amsmath,amssymb,amsfonts,lengthcheck,doublecolumn,superscriptaddress]{revtex4-2}
\usepackage{graphicx}
\usepackage{subfigure}
\usepackage{amsthm}
\usepackage{verbatim}
\usepackage{dcolumn}
\usepackage{bm}
\usepackage{epsf}
\usepackage{color}
\usepackage[colorlinks=true,citecolor=blue,linkcolor=blue,urlcolor=blue]{hyperref}%
\usepackage{xcolor}
\usepackage{dsfont}
\usepackage{tikz}
\usepackage{todonotes}
\usepackage{multirow}
\usepackage{xfrac}
\usepackage{makecell}

\newcommand{\pd}{\partial}

\newcommand{\bla}{bla\\bla\\bla\\bla\\bla}

\newcommand{\mc}[1]{\mathcal{#1}}

\newcommand{\mrm}[1]{\mathrm{#1}}

\DeclareMathAlphabet\mathbfcal{OMS}{cmsy}{b}{n}

\makeatletter
\newcommand{\currentfontsize}{\f@size pt}
\makeatother

\makeatletter
\newcommand\footnoteref[1]{\protected@xdef\@thefnmark{\ref{#1}}\@footnotemark}
\makeatother

\begin{document}

\title{Ideal heat engine cycles at maximal efficiency -- the ideal gas and beyond}

\author{Gregory Behrendt}
\affiliation{Department of Physics, University of Maryland, College Park, MD 20742 USA}
\affiliation{Department of Physics, University of Maryland, Baltimore County, Baltimore, MD 21250, USA}
\affiliation{Quantum Science Institute, University of Maryland, Baltimore County, MD 21250, USA}

\author{Sebastian Deffner}
\affiliation{Department of Physics, University of Maryland, Baltimore County, Baltimore, MD 21250, USA}
\affiliation{Quantum Science Institute, University of Maryland, Baltimore County, MD 21250, USA}
\affiliation{National Quantum Laboratory, College Park, MD 20740, USA}

\begin{abstract}
Given a particular heat engine cycle, what is the optimal working medium that results in the highest efficiency? While one might jump to the conclusion that it must surely be the ideal gas, the situation is actually more intricate. Starting with a general Helmholtz potential that depends polynomially on molar volume and temperature we derive exact expressions for the ideal Stirling, Otto, and Brayton cycles. We find that for the thermodynamic systems described by our ansatz for the Helmholtz potential the maximal efficiency is achieved, if the working medium is described by a fundamental relation linear in temperature. This includes the ideal gas, but also classical harmonic oscillators and phenomenological models of the rubber band.
\end{abstract}

\maketitle

\section{Introduction}

Computing the efficiency of ideal heat engine cycles is probably the most common problem in any undergraduate course in thermodynamics \cite{Callen1985}. In particular, the resulting efficiencies of ideal Otto and Stirling cycles have become general knowledge expected of anyone with a physics degree. What is maybe not always fully appreciated is the fact that especially in introductory, pedagogical treatments one typically assumes the engine to operate with an ideal gas as working medium. While this leads to simple formulas for the thermodynamic efficiencies, it hides the fact that for more general situations the efficiencies even of the ideal cycles can be significantly more involved.

This fact becomes particularly pertinent in the analysis of endoreversible heat engine cycles \cite{Deffner2018Entropy}. Here the term ``endoreversibility'' refers to thermodynamic processes of systems that follow successions of \emph{local equilibrium states}, but which are not in equilibrium with the ambient heat reservoirs \cite{Hoffmann1997}. For such cycles, one is the interested in the efficiency at maximal power, which for Carnot cycles is given by the Curzon-Ahlborn efficiency \cite{Curzon1975AJP},
\begin{equation}
\label{eq:CA}
\eta_\mrm{CA}=1-\sqrt{\frac{T_c}{T_h}}\,,
\end{equation}
where $T_c$ and $T_h$ are the temperatures of the cold and hot reservoirs, respectively. The very appealing simplicity of Eq.~\eqref{eq:CA} led to the somewhat erroneous assumption that the Curzon-Ahlborn formula universally describes the efficiency at maximal power, for which evidence was found for, e.g.,  endoreversible Otto engines with ideal gases \cite{Leff1987}, endoreversible Stirling cycles \cite{Erbay1997,Blanck1994Energy,Kaushik2000Energy}, Otto engines in the quasistatic limit \cite{Rezek2006,Esposito2010}, or for harmonic trap undergoing a quantum Otto cycle \cite{Abah2012,Rossnagel2014,Bonanca2018}.

However, at the latest and very succinctly in Ref.~\cite{Deffner2018Entropy} it was shown that the efficiency at maximal power depends on (i) the considered endoreversible cycle, and (ii) the fundamental relation of the working medium. This insight was further corroborated, e.g., in Ref.~\cite{Ferketic2023EPL} for cosmological engines and in Ref.~\cite{Behrendt2025} for plasma engines. Interestingly, these analyses highlighted that the particular choice of the working medium -- in comparison to the ideal gas -- can lead to either enhanced or decreased efficiency. See also Refs.~\cite{Abah2012,Kloc2019,Myers2020PRE,Myers2021Symmetry,Myers2021NJP,Myers2021PRXQ,Myers2022NJP,Myers2023Nanomat,Pena2023Entropy} for related findings. 

The somewhat natural question arises whether there is an optimal working medium that leads to the maximal efficiency. For endoreversible Otto cycles this question was already addressed in Ref.~\cite{Smith2020JNET}, however a comprehensive answer appears to be still lacking. It is interesting to note that the analysis of endoreversible cycles might actually obscure the pertinent issue. Note that endoreversible process are locally in equilibrium, and at least locally their efficiency is determined by the ideal expressions. At least for the scenarios considered in Ref.~\cite{Smith2020JNET} we observed that what appears locally optimal, also remains optimal at maximal power. In general this might not necessarily be the case, since maximizing the power can be a mathematically involved problem. Therefore, for the sake of simplicity and accessibility of the analysis, we restrict ourselves in the present analysis to the ideal engine cycles without maximizing the power.

In the present analysis, we thus ask a rather innocuously looking question: namely, for what working medium is the efficiency of an ideal heat engine cycle maximized? To this end, we will assume that the fundamental relation is of polynomial form, for which we can derive simple expression for the efficiencies of ideal Stirling, Otto, and Brayton cycles. We find that the maximal efficiencies are obtained for working mediums, whose caloric equations of state are linear in temperature -- this does include the ideal gas, but also describes, for instance, classical harmonic oscillators and rubber bands \cite{Callen1985}. We emphasize that similar question are often considered in engineering thermodynamics \cite{borgnakke2025fundamentals,Patel2022,Chai2025}, but to the very best of our knowledge our simple and pedagogical approach is unique and novel.

\section{Preliminaries: polynomial fundamental relation}

Motivated by Ref.~\cite{Smith2020JNET}, we consider a general thermodynamic system, whose Helmholtz potential can be written as
\begin{equation}
\label{eq:FR}
    F(T,v,N)=-f(n,m)\,N\,v^n\,T^m.
\end{equation}
Here, $N$ is the mole number, $v$ is the molar volume, and $T$ is the temperature. The allow range of parameters of the exponents $n$ and $m$ is determined by thermodynamic considerations of stability. A fundamental relation describes a stable system, if the thermodynamic potentials are convex functions of their extensive variables, and concave functions of their intensive variables \cite{Callen1985}. In the present case, we thus must have that
\begin{equation}
\frac{\pd^2 F}{\pd T^2}\bigg|_{v,N}\leq 0\quad\text{and}\quad \frac{\pd^2 F}{\pd v^2}\bigg|_{T,N}\geq 0\,.
\end{equation}
Thus, we the allow range of parameters of the exponents is
\begin{equation}
m\geq 1\quad\text{and}\quad n\leq 1\,.
\end{equation}
Finally, $f(n,m)$ is constant that depends on other physical properties of the thermodynamic system. Note that generally $f(n,m)$ has to be chosen to correctly model the thermodynamic properties of the considered system. This becomes particularly obvious when considering the resulting equations of state.

As usual \cite{Callen1985}, the corresponding mechanical equation of state given by
\begin{equation}
    P=-\frac{\partial F}{\partial V}\bigg|_T=f(n,m)\,n\,v^{n-1}\,T^m\,,
    \label{eq:pres}
\end{equation}
and for the caloric equation of state we obtain
\begin{equation}
    E=F-T\frac{\partial F}{\partial T}\bigg|_v=f(n,m)\,(m-1)\,N\,v^n\,T^m\,.
    \label{eq:EOS}
\end{equation}
The latter equation follows from explicitly expressing the entropy as a function of $T$ and $v$,
\begin{equation}
\label{eq:entropy}
    S=-\frac{\partial F}{\partial T}\bigg|_v=f(n,m)\,m\,N\,v^n\,T^{m-1}\,.
\end{equation}
Finally, to avoid clutter in the formulas, we will also write $f(n,m)=f$.

\subparagraph*{Example 1: photonic gas}

Before we continue, it is instructive to recall specific values of $n$ and $m$ for familiar examples. In particular, for $n=1$ and $m=4$ Eq.~\eqref{eq:FR} becomes
\begin{equation}
    F(T,v,N)=-f\,N\,v\,T^4.
\end{equation}
which we recognize as the fundamental relation of the photonic gas \cite{Callen1985} or the relativistic electron–positron–photon plasma \cite{Slattery1980PRA,Behrendt2025}.

\subparagraph*{Example 2: classical ideal gas}

As a second example, we briefly discuss how to obtain the ideal gas from Eq.~\eqref{eq:FR}. In its standard form \cite{Callen1985}, the Helmholtz potential of the classical ideal gas is given by
\begin{equation}
    F(T,v,N) = -Nk_B T \left[\ln\left(\frac{v}{\lambda^3}\right)+1\right]
\end{equation}
where $\lambda$ is the thermal wave length and $k_B$ Boltzmann's constant. This polynomial expression can be directly obtained from the polynomial formula in Eq.~\eqref{eq:FR}. To this end, recall that the natural logarithm can be expressed as a polynomial limit, namely \cite{Abramowitz1948}
\begin{equation}
    \lim_{n\to0}\left\{ \frac{1}{n}\left(x^n-1\right)\right\}=\ln(x).
\end{equation}
Thus, choosing $f\propto 1/n$ and $m=1$, the ideal gas is simply obtained from Eq.~\eqref{eq:FR} in the limit $n\to 0$ with $n/(m-1)\rightarrow\gamma-1$. As~always~\cite{Callen1985}, here $\gamma=c_P/c_v$ is the ratio of isobaric, $c_P$, and~isochoric, $c_v$, heat capacities.

\section{Ideal heat engine cycles}

Equipped with a general ansatz for the Helmholtz potential \eqref{eq:FR}, we can now continue to analyze ideal heat engine cycles. For pedagogical reasons, we will treat Stirling, Otto, and Brayton cycles separately. For each of these cycle, we will derive analytical expressions for the efficiency, which is generally defined as \cite{Callen1985}
\begin{equation}
\label{eq:efficiency}
    \eta=-\frac{W_\mrm{tot}}{Q_\mrm{in}}\leq 1- \frac{T_c}{T_h}\equiv \eta_\mrm{C}\,,
\end{equation}
and which has to be smaller than the Carnot efficiency, $\eta_\mrm{C}$.

\subsection{Stirling Cycle}

We begin with the Stirling cycle, which is comprised of two isothermal strokes and two isochoric strokes, namely:
\begin{itemize}
    \item[] $A\to B$: isothermal expansion
    \item[] $B\to C$: isochoric cooling 
    \item[] $C\to D$: isothermal compression
    \item[] $D\to A$: isochoric heating
\end{itemize}

To avoid confusion with the literature, especially from the engineering sciences \cite{borgnakke2025fundamentals}, we emphasize that we consider only Stirling cycles with perfect regeneration\footnote{Stirling engines with imperfect regeneration are a better description of real engines. However, in the present analysis we restrict ourselves to the simplest, most ideal scenario for the sake of clarity and simplicity.}. In this case, heat is only absorbed from the hot reservoir during $A\to B$, and that no work is performed during the isochoric strokes. Hence, the efficiency \eqref{eq:efficiency} can be written as
\begin{equation}
\label{eq:eta_S}
\eta_\mrm{S}=-\frac{W_{A\to B}+W_{C\to D}}{Q_{A\to B}}=-\frac{\Delta F_{A\to B}+\Delta F_{C\to D}}{Q_{A\to B}}\,,
\end{equation}
where we also used that in isothermal processes the work is given by the difference in Helmholtz potentials \cite{Callen1985}.

Using Eq.~\eqref{eq:FR}, we can now write
\begin{equation}
\Delta F_{A\to B}= -f\,N\,T_h^m\,\left(v^n_B-v^n_A\right)
\end{equation}
and
\begin{equation}
\Delta F_{C\to D}=-f\,N\,T_c^m\,\left(v^n_D-v^n_C\right)=f\,N\,T_c^m\,\left(v^n_B-v^n_A\right)
\end{equation}
where we further used $v_A=v_D$ and $v_B=v_C$. In addition, we have for the heat intake
\begin{equation}
Q_{A\to B}= T_h\,\Delta S_{A\to B}= f\,m\,N\,T_h^{m}\, \left(v^n_B-v_A^n\right)\,,
\end{equation}
which follows from Eq.~\eqref{eq:entropy}. Collecting terms and plugging into Eq.~\eqref{eq:eta_S}, the efficiency then simply becomes
\begin{equation}
  \label{eq:Stirling}
\eta_\mrm{S} =\frac{1}{m}\left[1-\left(\frac{T_c}{T_h}\right)^m\right]\,,
\end{equation}
which only depends on the ratio of cold and hot temperatures, $T_c/T_h$, and the exponent $m$.

\begin{figure}
    \centering
    \includegraphics[width=0.48\textwidth]{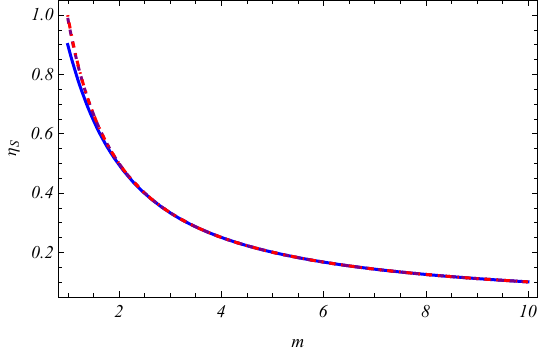}
    \caption{Stirling efficiency \eqref{eq:Stirling} as a function of $m$, for $T_c/T_h=10^{-1}$ (blue, solid line), $T_c/T_h=10^{-2}$ (purple, dashed line), and $T_c/T_h=10^{-3}$ (red, dot-dashed line)}
    \label{fig:stirling}
\end{figure}

Employing elementary calculus, it is easy to see that $\eta_\mrm{S}$ is a monotonically decreasing function of $m$ for any value of $T_c/T_h$. It assumes its maximum for $m=1$ for which $\eta_S$ becomes the Carnot efficiency, see also Fig.~\ref{fig:stirling} for an illustration. We emphasize that this ideal Stirling efficiency is independent of the exponent $n$. Thus, the maximal efficiency is attained for any caloric equations of state that are linear in temperatures, such as ideal gases, rubber bands, or classical harmonic oscillators \cite{Callen1985}.

\subsection{Otto Cycle}

As a second case, we treat the ideal Otto cycle, which is comprised of:
\begin{itemize}
    \item[] $A\to B$: isentropic compression
    \item[] $B\to C$:  isochoric heating
    \item[] $C\to D$:  isentropic expansion
    \item[] $D\to A$:  isochoric cooling
\end{itemize}

Again, no work is performed during the isochoric strokes, and no heat is exchanged during the isentropic strokes. Thus, we can now write \cite{Callen1985}
\begin{equation}
\label{eq:eta_O_1}
\eta_\mrm{O}=1-\frac{Q_\mrm{D\to A}}{Q_\mrm{B\to C}}\,.
\end{equation}
Moreover, we can also write,
\begin{equation}
Q_\mrm{B\to C}=\Delta E_\mrm{B\to C}\quad \text{and}\quad Q_\mrm{D\to A}=\Delta E_\mrm{D\to A}\,.
\end{equation}
The changes in internal energy are determined from the caloric equation of state \eqref{eq:EOS}, and we have
\begin{equation}
\Delta E_\mrm{B\to C}= f\,(m-1)\,N\,v_B^n\,(T_C^m-T_B^m)
\end{equation}
and
\begin{equation}
\Delta E_\mrm{D\to A}=f\,(m-1)\,N\,v_D^n\,(T_D^m-T_A^m)\,.
\end{equation}
Thus, the efficiency becomes
\begin{equation}
\label{eq:eta_O}
\eta_\mrm{O}=1-\frac{v_D^n}{v_B^n}\frac{T_D^m-T_A^m}{T_C^m-T_B^m}\,,
\end{equation}
which depends on the compression ratio $\kappa\equiv v_B/v_D$ and the four temperatures.

As usual, we can further simply $\eta_\mrm{O}$ by considering the isentropic strokes, namely $S_A=S_B$, and $S_C=S_D$. Using Eq.~\eqref{eq:entropy} we obtain
\begin{equation}
v_D^n\,T_A^{m-1}=v_B^n\,T_B^{m-1}\quad\text{and}\quad v_B^n\,T_C^{m-1}=v_D^n\,T_D^{m-1}\,,
\end{equation}
where we used that $v_A=v_D$ and $v_C=v_B$. In terms of the compression ratio, $\kappa$ we then also have
\begin{equation}
\frac{T_B^{m-1}}{T_A^{m-1}}=\kappa^{-n} \quad\text{and}\quad \frac{T_C^{m-1}}{T_D^{m-1}}=\kappa^{-n}\,
\end{equation}
from which we conclude that $T_B/T_A=T_C/T_D$. Thus, we can write
\begin{equation}
T_D^m-T_A^m=\left(T_C^m-T_B^m\right)\,\frac{T_A^m}{T_B^m}=\left(T_C^m-T_B^m\right)\,\kappa^{\frac{m n}{m-1}}\,,
\end{equation}
and the efficiency \eqref{eq:eta_O} finally becomes
\begin{equation}
\label{eq:Otto}
\eta_\mrm{O}=1-\kappa^{\frac{n}{m-1}}\,.
\end{equation}

\begin{figure}
    \centering
    \includegraphics[width=0.48\textwidth]{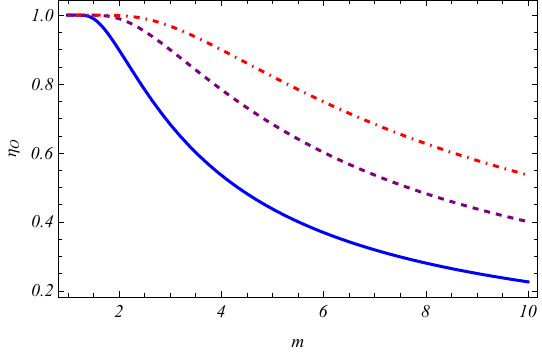}
    \caption{Otto efficiency \eqref{eq:Otto} as a function of $m$, for $n=1/2$ and $\kappa=10^{-1}$ (blue, solid line), $\kappa=10^{-2}$ (purple, dashed line), and $\kappa=10^{-3}$ (red, dot-dashed line)}
    \label{fig:otto}
\end{figure}

In contrast to the Stirling efficiency \eqref{eq:Stirling}, the Otto efficiency \eqref{eq:Otto} explicitly depends on both exponents, $n$ and $m$. Since the compression ratio $\kappa\leq 1$, the Otto efficiency is maximized for maximal values of the exponent $n/(m-1)$. For any finite value of $n$, this is achieved for again $m\to 1$, i.e., for caloric equations of state that are linear in temperature. Note that in principle the efficiency $\eta_\mrm{O}$ is also maximized in the limit $n\gg1$, which, however, would correspond to a highly unstable thermodynamic system. Our results for the Otto efficiency are sketched in Fig.~\ref{fig:otto}.

\subsection{Brayton Cycle}

As a third, and final case we consider the Brayton cycle. This cycle is identical to the Otto cycle, but replacing the isochoric with isobaric strokes. We have:
\begin{itemize}
    \item[] $A\to B$: isentropic compression 
    \item[] $B\to C$: isobaric heating
    \item[] $C\to D$:  isentropic expansion 
    \item[] $D\to A$:   isobaric cooling 
\end{itemize}

In complete analogy to the Otto cycle \eqref{eq:eta_O_1}, the efficiency is again given by,
\begin{equation}
\eta_\mrm{B}=1-\frac{Q_\mrm{D\to A}}{Q_\mrm{B\to C}}\,.
\end{equation}
Note, however, that the Brayton cycle is governed by isobaric strokes, which are best described in the enthalpy representation of the fundamental relation \cite{Callen1985}. To this end, we write
\begin{equation}
H=E+PV= f\left(m+n-1\right) N\,v^n\,T^m\,,
\end{equation}
where we used the caloric \eqref{eq:EOS} and mechanical \eqref{eq:pres} equations of state.

Therefore, we can also write
\begin{equation}
Q_\mrm{B\to C}=\Delta H_\mrm{B\to C}=f\left(m+n-1\right) N\,\left(v_C^n\,T_C^m-v_B^n\,T_B^m\right)
\end{equation}
and
\begin{equation}
Q_\mrm{D\to A}=\Delta H_\mrm{D\to A}=f\left(m+n-1\right) N\,\left(v_D^n\,T_D^m-v_A^n\,T_A^m\right)\,.
\end{equation}
Accordingly, the Brayton efficiency becomes
\begin{equation}
\label{eq:eta_B}
\eta_\mrm{B}=1-\frac{v_D^n\,T_D^m-v_A^n\,T_A^m}{v_C^n\,T_C^m-v_B^n\,T_B^m}\,.
\end{equation}

As before in the analysis of the Otto cycle, we now leverage the isentropic strokes. In particular, we again have
\begin{equation}
\label{eq:vols_bray}
v_A^n\,T_A^{m-1}=v_B^n\,T_B^{m-1}\quad\text{and}\quad v_C^n\,T_C^{m-1}=v_D^n\,T_D^{m-1}\,,
\end{equation}
which we now need to rewrite in terms of the pressure $P$. Again employing the mechanical equation of state \eqref{eq:pres}, Eq.~\eqref{eq:vols_bray} becomes
\begin{equation}
\label{eq:ps_bray}
\begin{split}
P_A^\frac{n}{n-1}\,T_A^{\frac{m}{1-n}-1}&=P_B^\frac{n}{n-1}\,T_B^{\frac{m}{1-n}-1}\\
P_B^\frac{n}{n-1}\,T_C^{\frac{m}{1-n}-1}&=P_A^\frac{n}{n-1}\,T_D^{\frac{m}{1-n}-1}\,,
\end{split}
\end{equation}
where we used that $P_B=P_C$ and $P_A=P_D$. 
Introducing the compressor ratio $\mc{C}=P_B/P_A$, the Brayton efficiency \eqref{eq:eta_B} can now be written as
\begin{equation}
\eta_\mrm{B}=1-\mc{C}^\frac{n}{1-n}\,\frac{T_D^{\frac{m}{1-n}}-T_A^{\frac{m}{1-n}}}{T_C^{\frac{m}{1-n}}-T_B^{\frac{m}{1-n}}}\,,
\end{equation}
which still depends on all four temperatures. 

The efficiency can be further simplified by inspecting Eq.~\eqref{eq:ps_bray}, which we re-write as
\begin{equation}
\left(\frac{T_A}{T_B}\right)^{\frac{1-m}{n}-1}=\mc{C}\quad\text{and}\quad \left(\frac{T_D}{T_C}\right)^{\frac{1-m}{n}-1}=\mc{C}\,,
\end{equation} 
and from which we again conclude that $T_A/T_B=T_D/T_C$. Now using the same ``trick'' as for the Otto cycle we obtain
\begin{equation}
\begin{split}
T_D^{\frac{m}{1-n}}-T_A^{\frac{m}{1-n}}&=\left(T_C^{\frac{m}{1-n}}-T_B^{\frac{m}{1-n}}\right)\,\frac{T_A^{\frac{m}{1-n}}}{T_B^{\frac{m}{1-n}}}\\
&=\left(T_C^{\frac{m}{1-n}}-T_B^{\frac{m}{1-n}}\right)\,\mc{C}^\frac{m n}{(n-1)(1-m-n)}\,,
\end{split}
\end{equation}
and thus finally
\begin{equation}
\label{eq:Brayton}
\eta_\mrm{B}=1-\mc{C}^\frac{n}{1-m-n}\,,
\end{equation}
which again depends on both exponents, $m$ and $n$.

\begin{figure}
    \centering
    \includegraphics[width=0.48\textwidth]{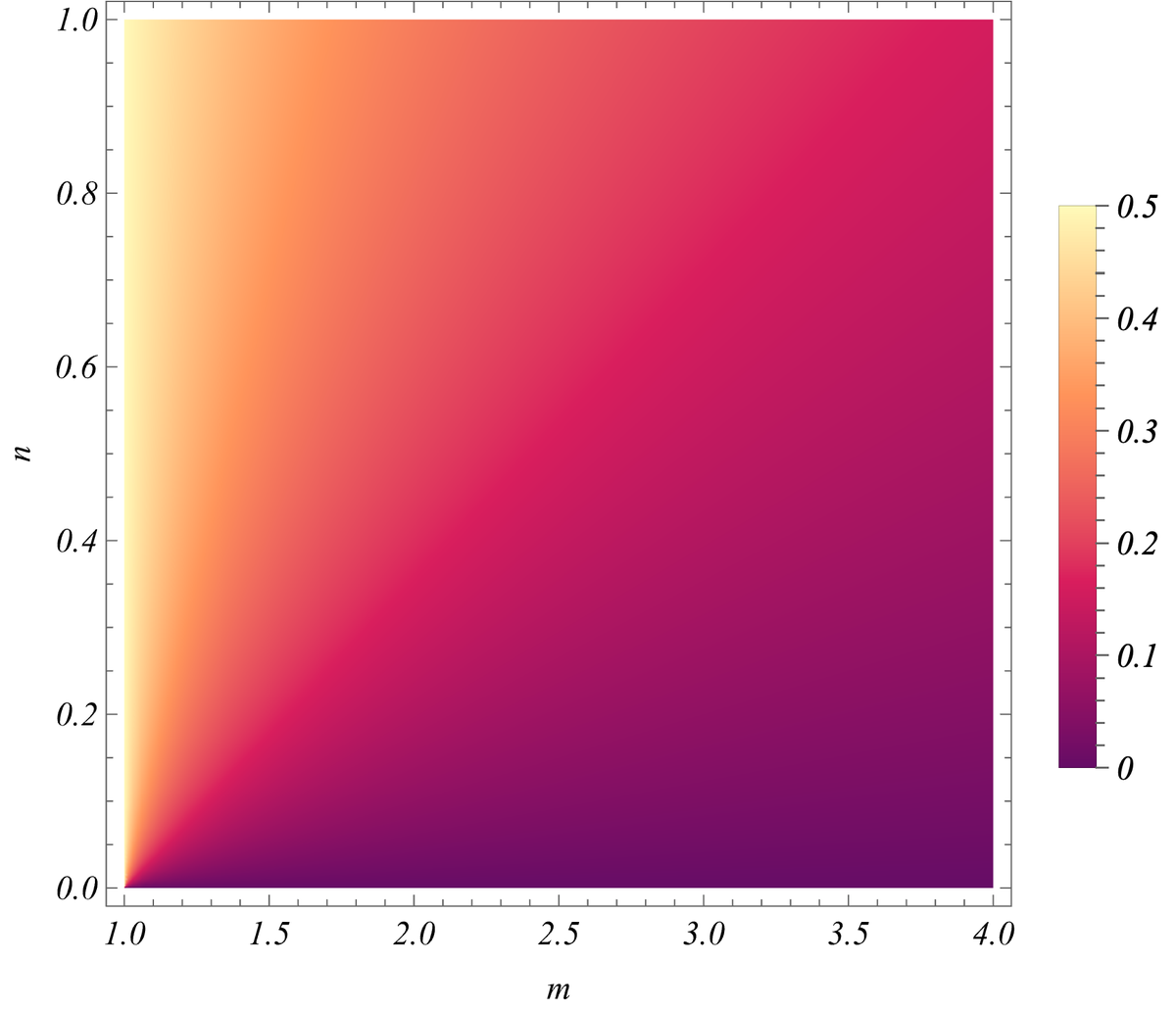}
    \caption{Brayton efficiency \eqref{eq:Brayton} as a function of $m$ and $n$ for $\mc{C}=2$. The range of parameters corresponds to fundamental relations \eqref{eq:FR} that describe thermodynamically stable systems.}
    \label{fig:brayton}
\end{figure}

In Fig.~\ref{fig:brayton} we illustrated the ideal Brayton efficiency \eqref{eq:Brayton} as a function of $n$ and $m$. We observe that the efficiency is again maximized in the non-physical, unstable limit $n\gg1$, but also again for $m=1$. Interestingly, in both limits we obtain
\begin{equation}
\lim_{m\to 1} \eta_\mrm{B}= 1- \mc{C}^{-1}\,.
\end{equation}
Thus we conclude that also the Brayton efficiency is maximized for working mediums whose fundamental relation is linear in temperature.

\section{Concluding remarks}

Starting from a general fundamental relation in Helmholtz representation, we derived closed-form expressions for the efficiencies of the ideal Stirling, Otto, and Brayton cycles. These analytical results enabled us to answer a rather innocuous-looking, but actually non-trivial question -- namely, which working medium results in the highest thermodynamic efficiency for a particular engine cycle. 

For all three cases we found that any thermodynamic system that is described by a caloric equation of state linear in temperature will give the highest possible efficiency. Interestingly, however, the Stirling efficiency is agnostic to the volume dependence of the Helmholtz potential, where we explicitly only considered ideal cycles with perfect regeneration. The Otto and Brayton efficiencies are slightly more involved, yet also their maximum is independent on the particular functional dependence of the fundamental relation on volume. 

More generally, our present work demonstrates that the heat engine efficiency of a given cycle is governed by the underlying structure of the working medium itself. This structure is captured by a generalized fundamental relation, revealing the thermodynamic properties that are most suited for maximal efficiency in a given engine cycle. Future work may include the analysis of more engine cycles, a more complex generalized fundamental relation, or explicit treatments of finite-time thermodynamic processes. The present analysis is entirely phrased in terms of thermodynamic notions. Hence, it will also be interesting to derive fundamental relations accounting for quantum effects that might lead to thermodynamics systems that exhibit performance beyond the polynomial equations of state discussed in this work.

\acknowledgments{G.B. thanks the Department of Physics at UMBC, which supported this research with a Donald N. Langenberg Undergraduate Research Award. S.D. acknowledges support from the John Templeton Foundation under Grant No. 63626.}

\bibliography{engine}

\begin{thebibliography}{30}%
\makeatletter
\providecommand \@ifxundefined [1]{%
 \@ifx{#1\undefined}
}%
\providecommand \@ifnum [1]{%
 \ifnum #1\expandafter \@firstoftwo
 \else \expandafter \@secondoftwo
 \fi
}%
\providecommand \@ifx [1]{%
 \ifx #1\expandafter \@firstoftwo
 \else \expandafter \@secondoftwo
 \fi
}%
\providecommand \natexlab [1]{#1}%
\providecommand \enquote  [1]{``#1''}%
\providecommand \bibnamefont  [1]{#1}%
\providecommand \bibfnamefont [1]{#1}%
\providecommand \citenamefont [1]{#1}%
\providecommand \href@noop [0]{\@secondoftwo}%
\providecommand \href [0]{\begingroup \@sanitize@url \@href}%
\providecommand \@href[1]{\@@startlink{#1}\@@href}%
\providecommand \@@href[1]{\endgroup#1\@@endlink}%
\providecommand \@sanitize@url [0]{\catcode `\\12\catcode `\$12\catcode
  `\&12\catcode `\#12\catcode `\^12\catcode `\_12\catcode `\%12\relax}%
\providecommand \@@startlink[1]{}%
\providecommand \@@endlink[0]{}%
\providecommand \url  [0]{\begingroup\@sanitize@url \@url }%
\providecommand \@url [1]{\endgroup\@href {#1}{\urlprefix }}%
\providecommand \urlprefix  [0]{URL }%
\providecommand \Eprint [0]{\href }%
\providecommand \doibase [0]{https://doi.org/}%
\providecommand \selectlanguage [0]{\@gobble}%
\providecommand \bibinfo  [0]{\@secondoftwo}%
\providecommand \bibfield  [0]{\@secondoftwo}%
\providecommand \translation [1]{[#1]}%
\providecommand \BibitemOpen [0]{}%
\providecommand \bibitemStop [0]{}%
\providecommand \bibitemNoStop [0]{.\EOS\space}%
\providecommand \EOS [0]{\spacefactor3000\relax}%
\providecommand \BibitemShut  [1]{\csname bibitem#1\endcsname}%
\let\auto@bib@innerbib\@empty
\bibitem [{\citenamefont {Callen}(1985)}]{Callen1985}%
  \BibitemOpen
  \bibfield  {author} {\bibinfo {author} {\bibfnamefont {H.}~\bibnamefont
  {Callen}},\ }\href@noop {} {\emph {\bibinfo {title} {Thermodynamics and an
  Introduction to Thermostastistics}}}\ (\bibinfo  {publisher} {Wiley},\
  \bibinfo {address} {New York, USA},\ \bibinfo {year} {1985})\BibitemShut
  {NoStop}%
\bibitem [{\citenamefont {Deffner}(2018)}]{Deffner2018Entropy}%
  \BibitemOpen
  \bibfield  {author} {\bibinfo {author} {\bibfnamefont {S.}~\bibnamefont
  {Deffner}},\ }\bibfield  {title} {\bibinfo {title} {Efficiency of harmonic
  quantum otto engines at maximal power},\ }\bibfield  {journal} {\bibinfo
  {journal} {Entropy}\ }\textbf {\bibinfo {volume} {20}},\ \href
  {https://doi.org/10.3390/e20110875} {10.3390/e20110875} (\bibinfo {year}
  {2018})\BibitemShut {NoStop}%
\bibitem [{\citenamefont {Hoffmann}\ \emph {et~al.}(1997)\citenamefont
  {Hoffmann}, \citenamefont {Burzler},\ and\ \citenamefont
  {Schubert}}]{Hoffmann1997}%
  \BibitemOpen
  \bibfield  {author} {\bibinfo {author} {\bibfnamefont {K.~H.}\ \bibnamefont
  {Hoffmann}}, \bibinfo {author} {\bibfnamefont {J.~M.}\ \bibnamefont
  {Burzler}},\ and\ \bibinfo {author} {\bibfnamefont {S.}~\bibnamefont
  {Schubert}},\ }\bibfield  {title} {\bibinfo {title} {Endoreversible
  thermodynamics},\ }\href {https://doi.org/10.1515/jnet.1997.22.4.311}
  {\bibfield  {journal} {\bibinfo  {journal} {J. Non-Equilib. Thermodyn.}\
  }\textbf {\bibinfo {volume} {22}},\ \bibinfo {pages} {311} (\bibinfo {year}
  {1997})}\BibitemShut {NoStop}%
\bibitem [{\citenamefont {Curzon}\ and\ \citenamefont
  {Ahlborn}(1975)}]{Curzon1975AJP}%
  \BibitemOpen
  \bibfield  {author} {\bibinfo {author} {\bibfnamefont {F.~L.}\ \bibnamefont
  {Curzon}}\ and\ \bibinfo {author} {\bibfnamefont {B.}~\bibnamefont
  {Ahlborn}},\ }\bibfield  {title} {\bibinfo {title} {{Efficiency of a {Carnot}
  engine at maximum power output}},\ }\href {https://doi.org/10.1119/1.10023}
  {\bibfield  {journal} {\bibinfo  {journal} {Am. J. Phys.}\ }\textbf {\bibinfo
  {volume} {43}},\ \bibinfo {pages} {22} (\bibinfo {year} {1975})}\BibitemShut
  {NoStop}%
\bibitem [{\citenamefont {Leff}(1987)}]{Leff1987}%
  \BibitemOpen
  \bibfield  {author} {\bibinfo {author} {\bibfnamefont {H.~S.}\ \bibnamefont
  {Leff}},\ }\bibfield  {title} {\bibinfo {title} {Thermal efficiency at
  maximum work output: {New} results for old heat engines},\ }\href
  {https://doi.org/10.1119/1.15071} {\bibfield  {journal} {\bibinfo  {journal}
  {Am. J. Phys.}\ }\textbf {\bibinfo {volume} {55}},\ \bibinfo {pages} {602}
  (\bibinfo {year} {1987})}\BibitemShut {NoStop}%
\bibitem [{\citenamefont {Erbay}\ and\ \citenamefont
  {Yavuz}(1997)}]{Erbay1997}%
  \BibitemOpen
  \bibfield  {author} {\bibinfo {author} {\bibfnamefont {L.~B.}\ \bibnamefont
  {Erbay}}\ and\ \bibinfo {author} {\bibfnamefont {H.}~\bibnamefont {Yavuz}},\
  }\bibfield  {title} {\bibinfo {title} {Analysis of the {Stirling} heat engine
  at maximum power conditions},\ }\href
  {https://doi.org/10.1016/S0360-5442(96)00159-4} {\bibfield  {journal}
  {\bibinfo  {journal} {Energy}\ }\textbf {\bibinfo {volume} {22}},\ \bibinfo
  {pages} {645} (\bibinfo {year} {1997})}\BibitemShut {NoStop}%
\bibitem [{\citenamefont {Blank}\ \emph {et~al.}(1994)\citenamefont {Blank},
  \citenamefont {Davis},\ and\ \citenamefont {Wu}}]{Blanck1994Energy}%
  \BibitemOpen
  \bibfield  {author} {\bibinfo {author} {\bibfnamefont {D.~A.}\ \bibnamefont
  {Blank}}, \bibinfo {author} {\bibfnamefont {G.~W.}\ \bibnamefont {Davis}},\
  and\ \bibinfo {author} {\bibfnamefont {C.}~\bibnamefont {Wu}},\ }\bibfield
  {title} {\bibinfo {title} {Power optimization of an endoreversible stirling
  cycle with regeneration},\ }\href
  {https://doi.org/https://doi.org/10.1016/0360-5442(94)90111-2} {\bibfield
  {journal} {\bibinfo  {journal} {Energy}\ }\textbf {\bibinfo {volume} {19}},\
  \bibinfo {pages} {125} (\bibinfo {year} {1994})}\BibitemShut {NoStop}%
\bibitem [{\citenamefont {Kaushik}\ and\ \citenamefont
  {Kumar}(2000)}]{Kaushik2000Energy}%
  \BibitemOpen
  \bibfield  {author} {\bibinfo {author} {\bibfnamefont {S.}~\bibnamefont
  {Kaushik}}\ and\ \bibinfo {author} {\bibfnamefont {S.}~\bibnamefont
  {Kumar}},\ }\bibfield  {title} {\bibinfo {title} {Finite time thermodynamic
  analysis of endoreversible stirling heat engine with regenerative losses},\
  }\href {https://doi.org/https://doi.org/10.1016/S0360-5442(00)00023-2}
  {\bibfield  {journal} {\bibinfo  {journal} {Energy}\ }\textbf {\bibinfo
  {volume} {25}},\ \bibinfo {pages} {989} (\bibinfo {year} {2000})}\BibitemShut
  {NoStop}%
\bibitem [{\citenamefont {Rezek}\ and\ \citenamefont
  {Kosloff}(2006)}]{Rezek2006}%
  \BibitemOpen
  \bibfield  {author} {\bibinfo {author} {\bibfnamefont {Y.}~\bibnamefont
  {Rezek}}\ and\ \bibinfo {author} {\bibfnamefont {R.}~\bibnamefont
  {Kosloff}},\ }\bibfield  {title} {\bibinfo {title} {Irreversible performance
  of a quantum harmonic heat engine},\ }\href
  {https://doi.org/10.1088/1367-2630/8/5/083} {\bibfield  {journal} {\bibinfo
  {journal} {New J. Phys.}\ }\textbf {\bibinfo {volume} {8}},\ \bibinfo {pages}
  {83} (\bibinfo {year} {2006})}\BibitemShut {NoStop}%
\bibitem [{\citenamefont {Esposito}\ \emph {et~al.}(2010)\citenamefont
  {Esposito}, \citenamefont {Kawai}, \citenamefont {Lindenberg},\ and\
  \citenamefont {Van~den Broeck}}]{Esposito2010}%
  \BibitemOpen
  \bibfield  {author} {\bibinfo {author} {\bibfnamefont {M.}~\bibnamefont
  {Esposito}}, \bibinfo {author} {\bibfnamefont {R.}~\bibnamefont {Kawai}},
  \bibinfo {author} {\bibfnamefont {K.}~\bibnamefont {Lindenberg}},\ and\
  \bibinfo {author} {\bibfnamefont {C.}~\bibnamefont {Van~den Broeck}},\
  }\bibfield  {title} {\bibinfo {title} {Efficiency at maximum power of
  low-dissipation {Carnot} engines},\ }\href
  {https://doi.org/10.1103/PhysRevLett.105.150603} {\bibfield  {journal}
  {\bibinfo  {journal} {Phys. Rev. Lett.}\ }\textbf {\bibinfo {volume} {105}},\
  \bibinfo {pages} {150603} (\bibinfo {year} {2010})}\BibitemShut {NoStop}%
\bibitem [{\citenamefont {Abah}\ \emph {et~al.}(2012)\citenamefont {Abah},
  \citenamefont {Ro\ss{}nagel}, \citenamefont {Jacob}, \citenamefont {Deffner},
  \citenamefont {Schmidt-Kaler}, \citenamefont {Singer},\ and\ \citenamefont
  {Lutz}}]{Abah2012}%
  \BibitemOpen
  \bibfield  {author} {\bibinfo {author} {\bibfnamefont {O.}~\bibnamefont
  {Abah}}, \bibinfo {author} {\bibfnamefont {J.}~\bibnamefont {Ro\ss{}nagel}},
  \bibinfo {author} {\bibfnamefont {G.}~\bibnamefont {Jacob}}, \bibinfo
  {author} {\bibfnamefont {S.}~\bibnamefont {Deffner}}, \bibinfo {author}
  {\bibfnamefont {F.}~\bibnamefont {Schmidt-Kaler}}, \bibinfo {author}
  {\bibfnamefont {K.}~\bibnamefont {Singer}},\ and\ \bibinfo {author}
  {\bibfnamefont {E.}~\bibnamefont {Lutz}},\ }\bibfield  {title} {\bibinfo
  {title} {Single-ion heat engine at maximum power},\ }\href
  {https://doi.org/10.1103/PhysRevLett.109.203006} {\bibfield  {journal}
  {\bibinfo  {journal} {Phys. Rev. Lett.}\ }\textbf {\bibinfo {volume} {109}},\
  \bibinfo {pages} {203006} (\bibinfo {year} {2012})}\BibitemShut {NoStop}%
\bibitem [{\citenamefont {Ro\ss{}nagel}\ \emph {et~al.}(2014)\citenamefont
  {Ro\ss{}nagel}, \citenamefont {Abah}, \citenamefont {Schmidt-Kaler},
  \citenamefont {Singer},\ and\ \citenamefont {Lutz}}]{Rossnagel2014}%
  \BibitemOpen
  \bibfield  {author} {\bibinfo {author} {\bibfnamefont {J.}~\bibnamefont
  {Ro\ss{}nagel}}, \bibinfo {author} {\bibfnamefont {O.}~\bibnamefont {Abah}},
  \bibinfo {author} {\bibfnamefont {F.}~\bibnamefont {Schmidt-Kaler}}, \bibinfo
  {author} {\bibfnamefont {K.}~\bibnamefont {Singer}},\ and\ \bibinfo {author}
  {\bibfnamefont {E.}~\bibnamefont {Lutz}},\ }\bibfield  {title} {\bibinfo
  {title} {Nanoscale heat engine beyond the {Carnot} limit},\ }\href
  {https://doi.org/10.1103/PhysRevLett.112.030602} {\bibfield  {journal}
  {\bibinfo  {journal} {Phys. Rev. Lett.}\ }\textbf {\bibinfo {volume} {112}},\
  \bibinfo {pages} {030602} (\bibinfo {year} {2014})}\BibitemShut {NoStop}%
\bibitem [{\citenamefont {Bonança}(2019)}]{Bonanca2018}%
  \BibitemOpen
  \bibfield  {author} {\bibinfo {author} {\bibfnamefont {M.~V.~S.}\
  \bibnamefont {Bonança}},\ }\bibfield  {title} {\bibinfo {title} {Approaching
  carnot efficiency at maximum power in linear response regime},\ }\href
  {https://doi.org/10.1088/1742-5468/ab4e92} {\bibfield  {journal} {\bibinfo
  {journal} {J. Stat. Mech.}\ }\textbf {\bibinfo {volume} {2019}},\ \bibinfo
  {pages} {123203} (\bibinfo {year} {2019})}\BibitemShut {NoStop}%
\bibitem [{\citenamefont {Ferketic}\ and\ \citenamefont
  {Deffner}(2023)}]{Ferketic2023EPL}%
  \BibitemOpen
  \bibfield  {author} {\bibinfo {author} {\bibfnamefont {E.~E.}\ \bibnamefont
  {Ferketic}}\ and\ \bibinfo {author} {\bibfnamefont {S.}~\bibnamefont
  {Deffner}},\ }\bibfield  {title} {\bibinfo {title} {Boosting thermodynamic
  performance by bending space-time},\ }\href
  {https://doi.org/10.1209/0295-5075/acad9c} {\bibfield  {journal} {\bibinfo
  {journal} {EPL (Europhys. Lett.)}\ }\textbf {\bibinfo {volume} {141}},\
  \bibinfo {pages} {19001} (\bibinfo {year} {2023})}\BibitemShut {NoStop}%
\bibitem [{\citenamefont {Behrendt}\ and\ \citenamefont
  {Deffner}(2025)}]{Behrendt2025}%
  \BibitemOpen
  \bibfield  {author} {\bibinfo {author} {\bibfnamefont {G.}~\bibnamefont
  {Behrendt}}\ and\ \bibinfo {author} {\bibfnamefont {S.}~\bibnamefont
  {Deffner}},\ }\bibfield  {title} {\bibinfo {title} {Endoreversible stirling
  cycles: Plasma engines at maximal power},\ }\bibfield  {journal} {\bibinfo
  {journal} {Entropy}\ }\textbf {\bibinfo {volume} {27}},\ \href
  {https://doi.org/10.3390/e27080807} {10.3390/e27080807} (\bibinfo {year}
  {2025})\BibitemShut {NoStop}%
\bibitem [{\citenamefont {Kloc}\ \emph {et~al.}(2019)\citenamefont {Kloc},
  \citenamefont {Cejnar},\ and\ \citenamefont {Schaller}}]{Kloc2019}%
  \BibitemOpen
  \bibfield  {author} {\bibinfo {author} {\bibfnamefont {M.}~\bibnamefont
  {Kloc}}, \bibinfo {author} {\bibfnamefont {P.}~\bibnamefont {Cejnar}},\ and\
  \bibinfo {author} {\bibfnamefont {G.}~\bibnamefont {Schaller}},\ }\bibfield
  {title} {\bibinfo {title} {Collective performance of a finite-time quantum
  otto cycle},\ }\href {https://doi.org/10.1103/PhysRevE.100.042126} {\bibfield
   {journal} {\bibinfo  {journal} {Phys. Rev. E}\ }\textbf {\bibinfo {volume}
  {100}},\ \bibinfo {pages} {042126} (\bibinfo {year} {2019})}\BibitemShut
  {NoStop}%
\bibitem [{\citenamefont {Myers}\ and\ \citenamefont
  {Deffner}(2020)}]{Myers2020PRE}%
  \BibitemOpen
  \bibfield  {author} {\bibinfo {author} {\bibfnamefont {N.~M.}\ \bibnamefont
  {Myers}}\ and\ \bibinfo {author} {\bibfnamefont {S.}~\bibnamefont
  {Deffner}},\ }\bibfield  {title} {\bibinfo {title} {Bosons outperform
  fermions: The thermodynamic advantage of symmetry},\ }\href
  {https://doi.org/10.1103/PhysRevE.101.012110} {\bibfield  {journal} {\bibinfo
   {journal} {Phys. Rev. E}\ }\textbf {\bibinfo {volume} {101}},\ \bibinfo
  {pages} {012110} (\bibinfo {year} {2020})}\BibitemShut {NoStop}%
\bibitem [{\citenamefont {Myers}\ \emph
  {et~al.}(2021{\natexlab{a}})\citenamefont {Myers}, \citenamefont {McCready},\
  and\ \citenamefont {Deffner}}]{Myers2021Symmetry}%
  \BibitemOpen
  \bibfield  {author} {\bibinfo {author} {\bibfnamefont {N.~M.}\ \bibnamefont
  {Myers}}, \bibinfo {author} {\bibfnamefont {J.}~\bibnamefont {McCready}},\
  and\ \bibinfo {author} {\bibfnamefont {S.}~\bibnamefont {Deffner}},\
  }\bibfield  {title} {\bibinfo {title} {Quantum heat engines with singular
  interactions},\ }\bibfield  {journal} {\bibinfo  {journal} {Symmetry}\
  }\textbf {\bibinfo {volume} {13}},\ \href
  {https://doi.org/10.3390/sym13060978} {10.3390/sym13060978} (\bibinfo {year}
  {2021}{\natexlab{a}})\BibitemShut {NoStop}%
\bibitem [{\citenamefont {Myers}\ \emph
  {et~al.}(2021{\natexlab{b}})\citenamefont {Myers}, \citenamefont {Abah},\
  and\ \citenamefont {Deffner}}]{Myers2021NJP}%
  \BibitemOpen
  \bibfield  {author} {\bibinfo {author} {\bibfnamefont {N.~M.}\ \bibnamefont
  {Myers}}, \bibinfo {author} {\bibfnamefont {O.}~\bibnamefont {Abah}},\ and\
  \bibinfo {author} {\bibfnamefont {S.}~\bibnamefont {Deffner}},\ }\bibfield
  {title} {\bibinfo {title} {Quantum otto engines at relativistic energies},\
  }\href {https://doi.org/10.1088/1367-2630/ac2756} {\bibfield  {journal}
  {\bibinfo  {journal} {New J. Phys.}\ }\textbf {\bibinfo {volume} {23}},\
  \bibinfo {pages} {105001} (\bibinfo {year} {2021}{\natexlab{b}})}\BibitemShut
  {NoStop}%
\bibitem [{\citenamefont {Myers}\ and\ \citenamefont
  {Deffner}(2021)}]{Myers2021PRXQ}%
  \BibitemOpen
  \bibfield  {author} {\bibinfo {author} {\bibfnamefont {N.~M.}\ \bibnamefont
  {Myers}}\ and\ \bibinfo {author} {\bibfnamefont {S.}~\bibnamefont
  {Deffner}},\ }\bibfield  {title} {\bibinfo {title} {Thermodynamics of
  statistical anyons},\ }\href {https://doi.org/10.1103/PRXQuantum.2.040312}
  {\bibfield  {journal} {\bibinfo  {journal} {PRX Quantum}\ }\textbf {\bibinfo
  {volume} {2}},\ \bibinfo {pages} {040312} (\bibinfo {year}
  {2021})}\BibitemShut {NoStop}%
\bibitem [{\citenamefont {Myers}\ \emph {et~al.}(2022)\citenamefont {Myers},
  \citenamefont {Peña}, \citenamefont {Negrete}, \citenamefont {Vargas},
  \citenamefont {De~Chiara},\ and\ \citenamefont {Deffner}}]{Myers2022NJP}%
  \BibitemOpen
  \bibfield  {author} {\bibinfo {author} {\bibfnamefont {N.~M.}\ \bibnamefont
  {Myers}}, \bibinfo {author} {\bibfnamefont {F.~J.}\ \bibnamefont {Peña}},
  \bibinfo {author} {\bibfnamefont {O.}~\bibnamefont {Negrete}}, \bibinfo
  {author} {\bibfnamefont {P.}~\bibnamefont {Vargas}}, \bibinfo {author}
  {\bibfnamefont {G.}~\bibnamefont {De~Chiara}},\ and\ \bibinfo {author}
  {\bibfnamefont {S.}~\bibnamefont {Deffner}},\ }\bibfield  {title} {\bibinfo
  {title} {Boosting engine performance with bose–einstein condensation},\
  }\href {https://doi.org/10.1088/1367-2630/ac47cc} {\bibfield  {journal}
  {\bibinfo  {journal} {New J. Phys.}\ }\textbf {\bibinfo {volume} {24}},\
  \bibinfo {pages} {025001} (\bibinfo {year} {2022})}\BibitemShut {NoStop}%
\bibitem [{\citenamefont {Myers}\ \emph {et~al.}(2023)\citenamefont {Myers},
  \citenamefont {Peña}, \citenamefont {Cortés},\ and\ \citenamefont
  {Vargas}}]{Myers2023Nanomat}%
  \BibitemOpen
  \bibfield  {author} {\bibinfo {author} {\bibfnamefont {N.~M.}\ \bibnamefont
  {Myers}}, \bibinfo {author} {\bibfnamefont {F.~J.}\ \bibnamefont {Peña}},
  \bibinfo {author} {\bibfnamefont {N.}~\bibnamefont {Cortés}},\ and\ \bibinfo
  {author} {\bibfnamefont {P.}~\bibnamefont {Vargas}},\ }\bibfield  {title}
  {\bibinfo {title} {Multilayer graphene as an endoreversible otto engine},\
  }\bibfield  {journal} {\bibinfo  {journal} {Nanomaterials}\ }\textbf
  {\bibinfo {volume} {13}},\ \href {https://doi.org/10.3390/nano13091548}
  {10.3390/nano13091548} (\bibinfo {year} {2023})\BibitemShut {NoStop}%
\bibitem [{\citenamefont {Peña}\ \emph {et~al.}(2023)\citenamefont {Peña},
  \citenamefont {Myers}, \citenamefont {Órdenes}, \citenamefont
  {Albarrán-Arriagada},\ and\ \citenamefont {Vargas}}]{Pena2023Entropy}%
  \BibitemOpen
  \bibfield  {author} {\bibinfo {author} {\bibfnamefont {F.~J.}\ \bibnamefont
  {Peña}}, \bibinfo {author} {\bibfnamefont {N.~M.}\ \bibnamefont {Myers}},
  \bibinfo {author} {\bibfnamefont {D.}~\bibnamefont {Órdenes}}, \bibinfo
  {author} {\bibfnamefont {F.}~\bibnamefont {Albarrán-Arriagada}},\ and\
  \bibinfo {author} {\bibfnamefont {P.}~\bibnamefont {Vargas}},\ }\bibfield
  {title} {\bibinfo {title} {Enhanced efficiency at maximum power in a
  fock–darwin model quantum dot engine},\ }\bibfield  {journal} {\bibinfo
  {journal} {Entropy}\ }\textbf {\bibinfo {volume} {25}},\ \href
  {https://doi.org/10.3390/e25030518} {10.3390/e25030518} (\bibinfo {year}
  {2023})\BibitemShut {NoStop}%
\bibitem [{\citenamefont {Smith}\ \emph {et~al.}(2020)\citenamefont {Smith},
  \citenamefont {Pal},\ and\ \citenamefont {Deffner}}]{Smith2020JNET}%
  \BibitemOpen
  \bibfield  {author} {\bibinfo {author} {\bibfnamefont {Z.}~\bibnamefont
  {Smith}}, \bibinfo {author} {\bibfnamefont {P.~S.}\ \bibnamefont {Pal}},\
  and\ \bibinfo {author} {\bibfnamefont {S.}~\bibnamefont {Deffner}},\
  }\bibfield  {title} {\bibinfo {title} {Endoreversible otto engines at maximal
  power},\ }\href {https://doi.org/doi:10.1515/jnet-2020-0039} {\bibfield
  {journal} {\bibinfo  {journal} {J. Non-Equilib. Thermodyn.}\ }\textbf
  {\bibinfo {volume} {45}},\ \bibinfo {pages} {305} (\bibinfo {year}
  {2020})}\BibitemShut {NoStop}%
\bibitem [{\citenamefont {Borgnakke}(2025)}]{borgnakke2025fundamentals}%
  \BibitemOpen
  \bibfield  {author} {\bibinfo {author} {\bibfnamefont {C.}~\bibnamefont
  {Borgnakke}},\ }\href@noop {} {\emph {\bibinfo {title} {Fundamentals of
  thermodynamics}}}\ (\bibinfo  {publisher} {John Wiley \& Sons},\ \bibinfo
  {year} {2025})\BibitemShut {NoStop}%
\bibitem [{\citenamefont {Patel}\ \emph {et~al.}(2022)\citenamefont {Patel},
  \citenamefont {Bass}, \citenamefont {Dukuze}, \citenamefont {Andrade},\ and\
  \citenamefont {Combs}}]{Patel2022}%
  \BibitemOpen
  \bibfield  {author} {\bibinfo {author} {\bibfnamefont {R.~C.}\ \bibnamefont
  {Patel}}, \bibinfo {author} {\bibfnamefont {D.~C.}\ \bibnamefont {Bass}},
  \bibinfo {author} {\bibfnamefont {G.~P.}\ \bibnamefont {Dukuze}}, \bibinfo
  {author} {\bibfnamefont {A.}~\bibnamefont {Andrade}},\ and\ \bibinfo {author}
  {\bibfnamefont {C.~S.}\ \bibnamefont {Combs}},\ }\bibfield  {title} {\bibinfo
  {title} {Analysis and development of a small-scale supercritical carbon
  dioxide (sco2) brayton cycle},\ }\bibfield  {journal} {\bibinfo  {journal}
  {Energies}\ }\textbf {\bibinfo {volume} {15}},\ \href
  {https://doi.org/10.3390/en15103580} {10.3390/en15103580} (\bibinfo {year}
  {2022})\BibitemShut {NoStop}%
\bibitem [{\citenamefont {Chai}\ and\ \citenamefont {Tassou}(2025)}]{Chai2025}%
  \BibitemOpen
  \bibfield  {author} {\bibinfo {author} {\bibfnamefont {L.}~\bibnamefont
  {Chai}}\ and\ \bibinfo {author} {\bibfnamefont {S.~A.}\ \bibnamefont
  {Tassou}},\ }\bibfield  {title} {\bibinfo {title} {A technology review of
  pumped thermal energy storage based on co2 cycles},\ }\href
  {https://doi.org/https://doi.org/10.1016/j.applthermaleng.2025.128586}
  {\bibfield  {journal} {\bibinfo  {journal} {Applied Thermal Engineering}\
  }\textbf {\bibinfo {volume} {281}},\ \bibinfo {pages} {128586} (\bibinfo
  {year} {2025})}\BibitemShut {NoStop}%
\bibitem [{\citenamefont {Slattery}\ \emph {et~al.}(1980)\citenamefont
  {Slattery}, \citenamefont {Doolen},\ and\ \citenamefont
  {DeWitt}}]{Slattery1980PRA}%
  \BibitemOpen
  \bibfield  {author} {\bibinfo {author} {\bibfnamefont {W.~L.}\ \bibnamefont
  {Slattery}}, \bibinfo {author} {\bibfnamefont {G.~D.}\ \bibnamefont
  {Doolen}},\ and\ \bibinfo {author} {\bibfnamefont {H.~E.}\ \bibnamefont
  {DeWitt}},\ }\bibfield  {title} {\bibinfo {title} {Improved equation of state
  for the classical one-component plasma},\ }\href
  {https://doi.org/10.1103/PhysRevA.21.2087} {\bibfield  {journal} {\bibinfo
  {journal} {Phys. Rev. A}\ }\textbf {\bibinfo {volume} {21}},\ \bibinfo
  {pages} {2087} (\bibinfo {year} {1980})}\BibitemShut {NoStop}%
\bibitem [{\citenamefont {Abramowitz}\ and\ \citenamefont
  {Stegun}(1948)}]{Abramowitz1948}%
  \BibitemOpen
  \bibfield  {author} {\bibinfo {author} {\bibfnamefont {M.}~\bibnamefont
  {Abramowitz}}\ and\ \bibinfo {author} {\bibfnamefont {I.~A.}\ \bibnamefont
  {Stegun}},\ }\href@noop {} {\emph {\bibinfo {title} {Handbook of mathematical
  functions with formulas, graphs, and mathematical tables}}},\ Vol.~\bibinfo
  {volume} {55}\ (\bibinfo  {publisher} {US Government printing office},\
  \bibinfo {year} {1948})\BibitemShut {NoStop}%
\bibitem [{Note1()}]{Note1}%
  \BibitemOpen
  \bibinfo {note} {Stirling engines with imperfect regeneration are a better
  description of real engines. However, in the present analysis we restrict
  ourselves to the simplest, most ideal scenario for the sake of clarity and
  simplicity.}\BibitemShut {Stop}%
\end{thebibliography}%

\end{document}